\documentclass{article}

\usepackage{arxiv}

\usepackage{amssymb}
\usepackage{pifont}
\newcommand{\cmark}{\ding{51}}%
\newcommand{\xmark}{\ding{55}}%

\usepackage[utf8]{inputenc} 
\usepackage[T1]{fontenc}    
\usepackage{comment}
\usepackage{hyperref}       
\usepackage{url}            
\usepackage{booktabs}       
\usepackage{amsfonts}       
\usepackage{nicefrac}       
\usepackage{microtype}      
\usepackage{lipsum}		
\usepackage{graphicx}
\usepackage{todonotes}
\usepackage{natbib}
\usepackage{doi}
\usepackage{subcaption}
\usepackage{amsmath}
\usepackage{booktabs} 

\title{Exploring the Potential of Data-Driven Spatial Audio Enhancement Using a Single-Channel Model}

\date{} 

\author{ \href{https://orcid.org/0000-0002-3989-7105}{\includegraphics[scale=0.06]{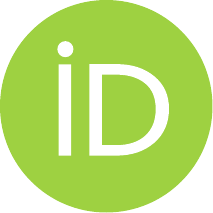}\hspace{1mm}Arthur N. dos Santos}
  \\
	Faculdade de Engenharia Elétrica e de Computação\\
	Universidade Estadual de Campinas\\
	Campinas, Brazil \\
	\texttt{a264372@dac.unicamp.br} \\
	\And
	\href{https://orcid.org/0000-0002-2246-4450}{\includegraphics[scale=0.06]{orcid.pdf}\hspace{1mm}Bruno S. Masiero} \\
	Faculdade de Engenharia Elétrica e de Computação\\
	Universidade Estadual de Campinas\\
	Campinas, Brazil \\
	\texttt{masiero@unicamp.br} \\
	\And
	\href{https://orcid.org/0009-0002-9822-9761}{\includegraphics[scale=0.06]{orcid.pdf}\hspace{1mm}Túlio C. L. Mateus} \\
	Faculdade de Engenharia Acústica\\
	Universidade Federal de Santa Maria\\
	Santa Maria, Brazil \\
	\texttt{tulio.chiodi@eac.ufsm.br } \\
}

\renewcommand{\shorttitle}{\raggedright Exploring the Potential of Data-Driven Spatial Audio Enhancement Using a Single-Channel Model}

\hypersetup{
pdftitle={Exploring the Potential of Data-Driven Spatial Audio Enhancement Using a Single-Channel Model},
pdfauthor={Arthur N. dos Santos, Bruno S. Masiero, Túlio C. L. Mateus},
pdfkeywords={single-channel, multi-channel, speech enhancement, dereverberation, spatial audio},
}

\begin{document}
\maketitle

\begin{abstract}
One key aspect differentiating data-driven single- and multi-channel speech enhancement and dereverberation methods is that both the problem formulation and complexity of the solutions are considerably more challenging in the latter case. Additionally, with limited computational resources, it is cumbersome to train models that require the management of larger datasets or those with more complex designs. In this scenario, an unverified hypothesis that single-channel methods can be adapted to multi-channel scenarios simply by processing each channel independently holds significant implications, boosting compatibility between sound scene capture and system input-output formats, while also allowing modern research to focus on other challenging aspects, such as full-bandwidth audio enhancement, competitive noise suppression, and unsupervised learning. This study verifies this hypothesis by comparing the enhancement promoted by a basic single-channel speech enhancement and dereverberation model with two other multi-channel models tailored to separate clean speech from noisy 3D mixes. A direction of arrival estimation model was used to objectively evaluate its capacity to preserve spatial information by comparing the output signals with ground-truth coordinate values. Consequently, a trade-off arises between preserving spatial information with a more straightforward single-channel solution at the cost of obtaining lower gains in intelligibility scores.
\end{abstract}


\section{Introduction}\label{intro}

Data-driven Speech Enhancement (SE) and dereverberation methods are pivotal in improving the quality and intelligibility of speech signals in various acoustic scenarios. These often rely on statistical methods, signal processing, and feature engineering to establish nonlinear complex maps between clean and noisy signals~\cite{lyon2017human}. Inspired by the well-known ``cocktail party effect''~\cite{Cherry1957}, recent advancements have incorporated microphone array processing by leveraging different geometries to capture spatial information. Consequently, multi-channel SE and dereverberation have garnered significant attention for applications in augmented, virtual, and mixed realities~\cite{refId0}.

This surge of interest on spatial audio enhancement has given rise to specialized data challenges, such as Learning 3D Audio Sources (L3DAS) 2021\footnote{\url{https://www.l3das.com/mlsp2021/}}, 2022\footnote{\url{https://www.l3das.com/icassp2022/index.html}} and 2023\footnote{\url{https://www.l3das.com/icassp2023/}}. These challenges typically provide access to baseline models and custom training datasets, thereby stimulating research on this topic. The Filter and Sum Network~(FaSNet)~\cite{9003849}, which is the baseline model for Task 1 of L3DAS21~\cite{9596248}, combines Normalized Cross-Correlation (NCC), Temporal Convolution Network (TCN) layers, and beamforming filters to separate desired audio components from background noise or interference. Notably, a multi-channel U-net with a neural beamformer (MMUB)~\cite{ren21} achieved the challenge's top rank, outperforming FaSNet on both Word Error Rate (WER) and Short-Time Objective Intelligibility~(STOI) metrics, and became the new baseline model for Task 1 of L3DAS22~\cite{9746872} and L3DAS23.

However, data-driven SE and dereverberation methods usually encounter various obstacles (especially for the multi-channel case), such as effectively capturing and processing acoustic scenarios, managing large datasets with limited computational resources, addressing model design complexity, and selecting suitable evaluation metrics. A recent survey~\citet{dosSantos2022} showed that only a fraction of articles published in 2021 on data-driven speech and audio enhancement methods explicitly claimed to be suitable for multi-channel cases (23\%), whereas a similar proportion referred to single-channel cases (24\%). Although a few others even explored more specific cases, such as binaural, ambisonics, and stereo signals (9\%), the vast majority (44\%) did not specify their intended use case scenario.

It is not only important to question why so many authors omitted this information, as it is also worth noting that many of the so-called multi-channel solutions perform a signal down-mix, i.e., producing an output signal that discards spatial information. While focusing on single-channel solutions may be justified in contexts such as communications, in which edge devices predominantly feature single microphones for capturing voice signals, in other applications such as hearing aid technology, multi-channel solutions may offer even more benefits for end users.

However, the loose hypothesis that single-channel methods can be adapted to multi-channel scenarios simply by processing each channel independently~\cite{dosSantos2022} holds significant implications for the compatibility of audio formats and data-driven speech and audio enhancement models. If confirmed, this hypothesis could revolutionize the field, allowing for more versatile applications and advancements.

The remainder of this paper is organized as follows: Section \ref{materials} describes three different SE and dereverberation models and another for Direction of Arrival (DOA) estimation, all used as baseline materials and methods in this study. In Section \ref{methods}, the problem formulation is presented, followed by the solutions offered by each model. The experimental results are presented in Section \ref{exp} and are discussed in Section \ref{disc}. Finally, the conclusions are presented in Section \ref{conc}.

\section{Materials and Methods}\label{materials}

In this study, a set of pre-trained data-driven models was used. The following subsections describe their primary purpose, architecture, and training datasets used.

\subsection{Multi-Channel SE Models}\label{mcse}

All multi-channel SE models are linked to the previous editions of the L3DAS challenge. FaSNet was the baseline model for Task 1 of L3DAS21 and was later replaced by MMUB in subsequent challenge editions.

\subsubsection{Filter and Sum Network (FaSNet)}\label{FaSNet}

FaSNet~\citep{9003849} is a Deep Learning (DL) architecture primarily intended for Source Separation (SS) that utilizes a series of Convolutional Recurrent Neural Network (CRNN) modules. These modules incorporate convolutional layers for local feature extraction and recurrent layers to model temporal dependencies. In addition, each module integrates a time-domain filter to capture source-specific information, and a summing operation to combine filtered signals, producing separate sources. Because FaSNet comprises stacked modules, the output of one module serves as the input to the next module, allowing progressive refinement of separation. This process involves computing an NCC metric between the time windows of a reference channel and the remaining channels, pooling the metric, passing it through a TCN, and feeding it into the beamforming filter learner. Subsequently, the NCC metric is computed between the previously denoised time windows and the remaining channels. Ultimately, each channel is individually denoised and summed, resulting in a final beamformed signal. Consequently, this model demonstrates the capability of enhancing noisy signals by effectively separating the desired component from background noise or interfering sources.
 
This study uses a pre-trained model\footnote{\url{https://github.com/l3das/L3DAS21}} of FaSNet. The training dataset comprises clean speech samples from LibriSpeech\footnote{\url{https://www.openslr.org/12}} (up to 10~s in duration), 12 transient (computer keyboard, drawer, cupboard, finger snapping, keys jangling, knock, laughter, scissors, telephone, writing, chink and clink, printer) and 4 continuous (alarm, crackle, mechanical fan and microwave oven) classes of noise samples from FSD50K\footnote{\url{https://zenodo.org/record/4060432\#.ZG0e5HbMJPY}}, and 252 Room Impulse Response (RIR) positions collected in an office-like environment. The model’s input is tailored for first-order ambisonics (FOA) signals, whereas its output is a monophonic signal.

\subsubsection{Multi-Channel U-net with Neural Beamformer (MMUB)}\label{MMUB}

MMUB~\cite{ren21} is a model initially submitted to L3DAS21, combining a Multiple-Input (MI) U-Net architecture with a Single-Output (SO) neural beamformer, referred to as ``MIMO U-net Beamforming'' in both the L3DAS22 and L3DAS23 documentations, despite being a MISO system. It uses FOA signals as the input and employs Short-Time Fourier Transform (STFT) as a feature extraction method. The resulting time-frequency representation undergoes encoding and decoding through the contracting and expanding paths of a U-Net. This enables dynamic adjustment of the neural beamformer weights based on the learned parameters. The architecture encompasses encoder blocks for high-level feature extraction, decoder blocks for reconstructing the input feature size, and skip connections for concatenating each layer in the encoder blocks with the corresponding layers in the decoder blocks.

In this study, a pre-trained model\footnote{\url{https://github.com/l3das/L3DAS22}} of MMUB is used. The training dataset is the same as that described in Section \ref{FaSNet}.

\subsection{Single-Channel SE Model (Fully-Convolutional U-net)}\label{FCU-net}

This Fully Convolutional (FC) U-net~\cite{ernst2018speech} follows the original work of~\cite{10.1007/978-3-319-24574-4_28}, which is a typical 2D Convolutional Neural Network (CNN) architecture with a contracting path to capture the context and a symmetric expanding path that enables precise localization. While each downsampling increases the number of feature channels, the expansive path consists of an upsampling of the feature map, decreasing the number of feature channels, which allows the network to propagate context information to higher-resolution layers. Consequently, the expansive path is approximately symmetric to the contracting path, yielding a (sometimes inverted) U-shaped architecture.

Although the authors in~\cite{10.1007/978-3-319-24574-4_28} developed this model for biomedical image segmentation, in~\cite{ernst2018speech} it was indicated that when considering a feature extractor based on STFT, this architecture can be helpful for SE, especially in dereverberation tasks. Although some authors claim that conversion to and from the time-frequency domain can be cumbersome, using spectrograms as input data effectively mitigates reverberation effects by capturing important spectral patterns of speech.

In this study, a pre-trained model\footnote{\url{https://www.mathworks.com/help/audio/ug/dereverberate-speech-using-deep-learning-networks.html}} of this FC U-net is used. The training dataset comprises noisy\footnote{\url{https://datashare.ed.ac.uk/handle/10283/2791}} and reverberant\footnote{\url{https://datashare.ed.ac.uk/handle/10283/2031}} speech samples from the VCTK dataset.

\subsection{DOA Estimation Model (Sound Event Localization and Detection Network)}\label{SELDnet}

Sound Event Localization and Detection Network (SELDnet)~\cite{8567942} is a Multi-Task Learning (MTL) model that consists of two independently trained CRNNs, one for Sound Event Detection (SED) and the other for DOA estimation.

The SED model uses log-magnitude STFT as predictors to the system. The network is then implemented in two stages: CNN layers and Gated Recurrent Units (GRU). A custom reshaping layer is used to recast the output of the CNN and pass it to the input of the RNN model. The final activation layer uses a sigmoid function.

The DOA estimation model uses a generalized cross-correlation phase transform (GCC-PHAT) as the system predictor. The network itself is similar to that of the SED model. The key difference is the size of the input layer and final activation layer, which uses a hyperbolic tangent function.

In this study, a pre-trained model\footnote{\url{https://github.com/l3das/L3DAS21}} of SELDnet is used, which was the baseline model for Task 2 of L3DAS21. The training dataset is similar to that described in Section \ref{FaSNet}.

\section{Experimental Setup}\label{methods}

In all editions of L3DAS, clean speech samples were augmented to obtain synthetic 3D acoustic scenes containing a randomly placed speaker, typical background noise, and reverberation of an office-like environment. In the following subsections, the problem formulation and solutions offered by each model described in Section \ref{materials} are presented.

\subsection{Problem Formulation}\label{problem}

Let $x(t)$ denote a clean monophonic speech sample and $\mathbf{H}(t)$ be the measured Room Impulse Response (RIR) in B-format:
\begin{equation}
    \mathbf{H}(t) = \left[ h(t)_W \ h(t)_X \ h(t)_Y \ h(t)_Z \right]^{T}\textrm{,}
\end{equation}
in which the sub-indices represent the corresponding $W$, $X$, $Y$ and $Z$ channels. To generate a reverberant spatial sound scene, $x(t)$ can be convolved with $\mathbf{H}(t)$:
\begin{equation}
    \mathbf{S}(t) = x(t) * \mathbf{H}(t)\textrm{,}
\end{equation}
such that $\mathbf{S}(t)$ represents the reverberant spatial sound scene. Similarly, a monophonic noise sample $n(t)$ can be convolved with another RIR to obtain $\mathbf{N}(t)$, i.e., a reverberant spatial noise sample in B-format. Hence, the total mix can be written as:
\begin{equation}
    \mathbf{Y}(t) = \mathbf{S}(t) + \mathbf{N}(t),
\end{equation}
in which $\mathbf{Y}(t)$ represents a noisy reverberant spatial sound scene.

\subsection{Solutions Under Consideration}\label{MCSE}

\begin{enumerate}
    \item \textbf{FaSNet}: When considering the 3D SE performed by FaSNet, which is a black-box-type system working in time domain, one should consider $\mathbf{Y}(t)$ as input, and $\hat{x}(t)$ as an estimation of the clean monophonic signal of interest, as free from additive and convolutive noise as possible;
    \item \textbf{MMUB}: Moreover, in the 3D SE performed by MMUB, it produces an output $\hat{X}(\tau, f)$ which is as an estimation of the clean monophonic speech matrix computed from a noisy reverberant spatial sound scene matrix $\mathbf{Y}(\tau, f)$, in which $\tau$ and $f$ are the time and frequency indices in the STFT domain, respectively. Because the MMUB computes real and imaginary masks, conversion to and from the time-frequency domain is a complex operation. 
    \item \textbf{FC U-net}: The pre-trained FC U-net is also a black-box-type system, which produces an output $\hat{X}(\tau, f)$ that is an estimation of the clean monophonic speech matrix computed from a noisy reverberant matrix $Y(\tau, f)$. Thus, to obtain $\hat{x}(t)$ from $y(t)$, conversion to and from the time-frequency domain is necessary. However, because FC U-net only estimates the magnitude of $\hat{X}(\tau, f)$, the original phase of $Y(\tau, f)$ is used in the inverse-STFT operation. Hence, it did not affect the Inter-Channel Phase Difference (ICPD), and no normalization was applied between the input and output channels.
\end{enumerate}

\subsection{Audio Example}

\begin{figure}[ht!]
\centering
\includegraphics[width=\textwidth]{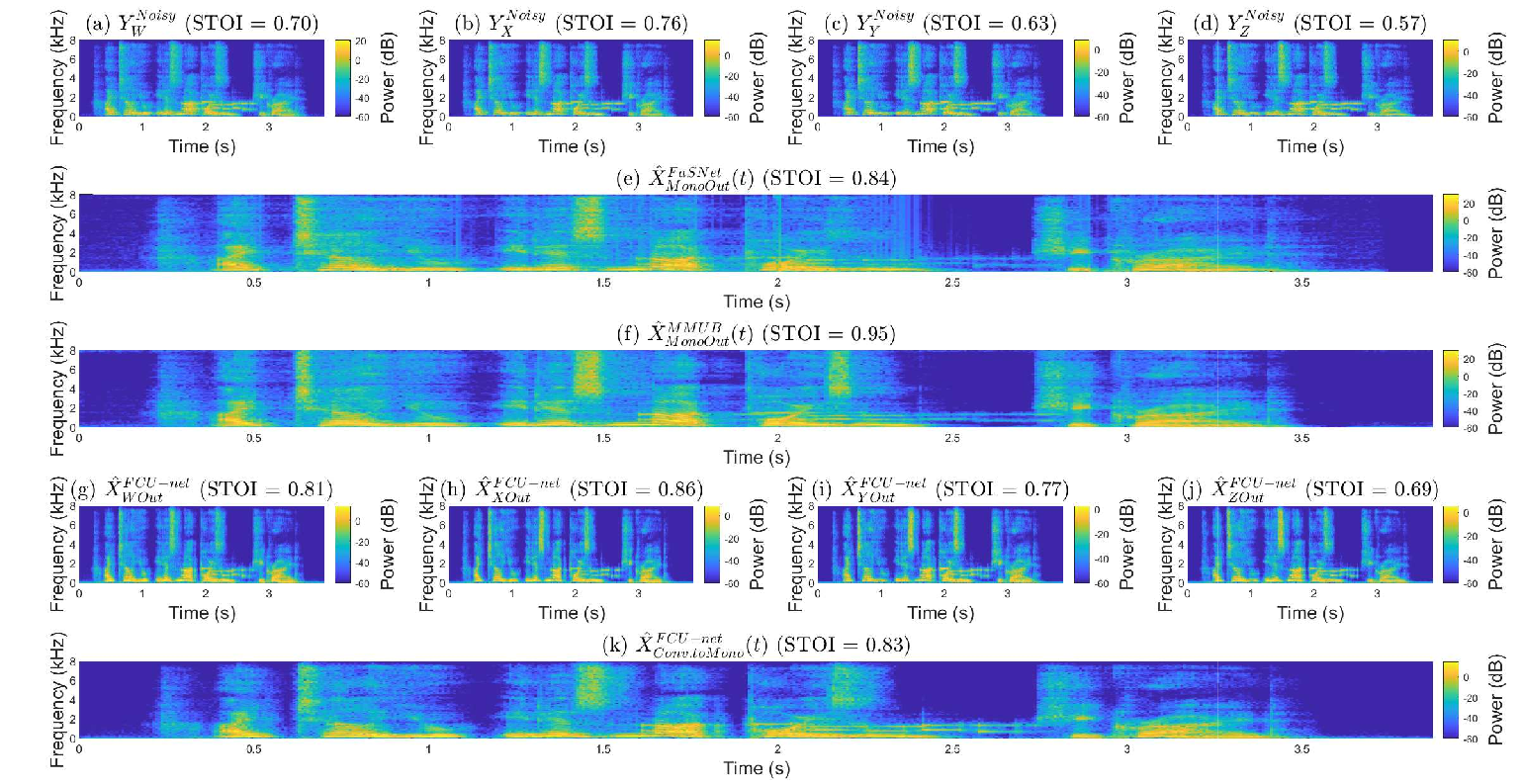}
\caption{The 4--channels of a noisy reverberant spatial sound scene (a--d), followed by the clean monophonic speech estimated by FaSNet (e) and MMUB (f), the enhancement promoted by the FC U-net for each channel independently (g--j), and a conversion to Mono format in the DOA detected by SELDnet. STOI scores are computed with reference to the clean monophonic speech signal $x(t)$.}
\label{fig:fasnetenh}
\end{figure}

Consider the four channels of a noisy reverberant spatial sound scene $\mathbf{Y}(t)$, as illustrated in Figure~\ref{fig:fasnetenh}(a--d). Consequently, Figure \ref{fig:fasnetenh}(e) shows a spectrogram of the FasNet estimation of the clean monophonic speech $\hat{x}(t)$, and \ref{fig:fasnetenh}(f) represents $\hat{X}(\tau, f)$ as outputted by MMUB. Subsequently, Figure~\ref{fig:fasnetenh}(g--j) shows the enhancement promoted by FC U-net for each of the four channels independently. STOI scores (cf. \cite{taal2010short}) are shown for all signals with reference to $x(t)$. Using SELDnet to estimate the DOA of the desired sound source from the enhanced sound scene yielded $153.10^\circ$ of azimuth and $12.71^\circ$ of elevation. By converting this enhanced sound scene to the Mono format, in the direction $(153.10^\circ, 12.71^\circ)$, a fairer comparison with the other methods is provided (k). To listen to these samples, the reader in referred to this link\footnote{\url{http://tinyurl.com/yc6t4fc3}}.

In this example, both FaSNet and MMUB were able to mask additive noise (a telephone at approximately 2.5~s). The MMUB achieves a higher STOI score owing to better mitigation of reverberation effects. However, both multi-channel solutions completely discard spatial information owing to the beamforming filters.

A comparison of the SELDnet estimation of the DOA of the desired sound source from the sound scene version enhanced by the FC U-net and its ground-truth coordinate values returned an azimuthal separation of $-0.48^\circ$ and an angular separation of $0.25^\circ$ in elevation. Hence, by not altering both the Inter-Channel Level Difference (ICLD) and ICPD, it was possible to preserve the spatial information with minimum error.

\begin{figure}[ht!]
\centering
\includegraphics[width=0.75\textwidth]{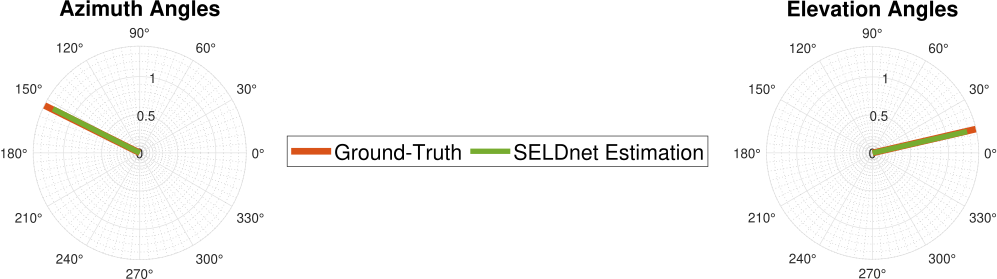}
\caption{Polar plots for the associated azimuth and elevation angles of the estimated desired sound source's DOA and ground-truth coordinate values.}
\label{fig:angles}
\end{figure}

Figure \ref{fig:angles} illustrates the polar plots for the associated azimuth and elevation angles of the SELDnet estimation and the ground-truth coordinate values of the DOA of the desired sound source.

\subsection{Summary}

Table \ref{tab:my-table} summarizes the advantages and disadvantages of each SE model considered in the audio example. Although the FC U-net does not incorporate array processing, i.e., it does not leverage multiple-input data to produce an output, nor does it tackle additive noise, it is the only model capable of preserving spatial information.

\begin{table}[ht!]
\centering
\caption{Comparison of data-driven SE models considered in this example.}
\label{tab:my-table}
\resizebox{0.80\columnwidth}{!}{%
\begin{tabular}{@{}ccccccc@{}}
\toprule
 & \textbf{} & \textbf{System} & \textbf{Masks} & \textbf{Masks} & \textbf{Incorporates} & \textbf{Preserves} \\
 & \textbf{Working} & \textbf{I/O} & \textbf{Convolutive} & \textbf{Additive} & \textbf{Array} & \textbf{Spatial} \\
 & \textbf{Domain} & \textbf{Format} & \textbf{Noise?} & \textbf{Noise?} & \textbf{Processing?} & \textbf{Information?} \\
\midrule
\textbf{FC U-net} & Magnitude STFT & SISO & \cmark & \xmark & \xmark & \cmark \\
\textbf{FaSNet} & Time & MISO & \cmark & \cmark & \cmark & \xmark \\
\textbf{MMUB} & Complex STFT & MISO & \cmark & \cmark & \cmark & \xmark \\
\bottomrule
\end{tabular}%
}
\end{table}

\section{Results}\label{exp}

To verify whether the experimental setup described in Section \ref{methods} would hold for a variety of 3D sound scenes, the test set for Task 1 of L3DAS22 was used in this study. It is comprised of 2.189 samples, in which the predictors consist of two sets of B-format FOA recordings (4--channels each), namely ``Mic A'' and ``Mic B'', released as .wav files, with a sampling rate of 16~kHz and a resolution of 16~bit. In this setting, Mic A is placed at the center of an office-like room, with dimensions 6~m (length) by 5~m (width) by 3~m (height), while Mic B is horizontally separated from Mic A by 20~cm. Both microphones are at a height of 1.3~m.
The targets are clean monophonic speech samples and the ground-truth $X$, $Y$, and $Z$ coordinates of each sound event's DOA, together with their respective spatial distances~\cite{9746872}.

\subsection{MISO Solutions}

\begin{figure}[ht!]
\centering
\includegraphics[width=0.75\textwidth]{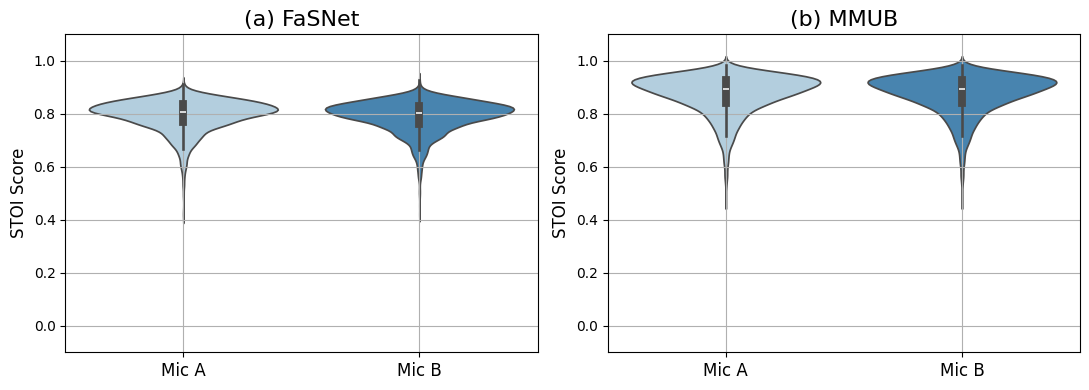}
\caption{STOI scores for the enhancement promoted by (a) FaSNet and (b) MMUB, using the L3DAS22 test set.}
\label{fig:fnmmub}
\end{figure}

Figure \ref{fig:fnmmub} illustrates the violin plots for the STOI scores obtained using the multi-channel methods described in Section \ref{mcse} to enhance the number of samples in the test set. Because both models are MISO, only one output per microphone is shown.

\subsection{SISO Solution}

\begin{figure}[ht!]
\centering
\includegraphics[width=0.75\textwidth]{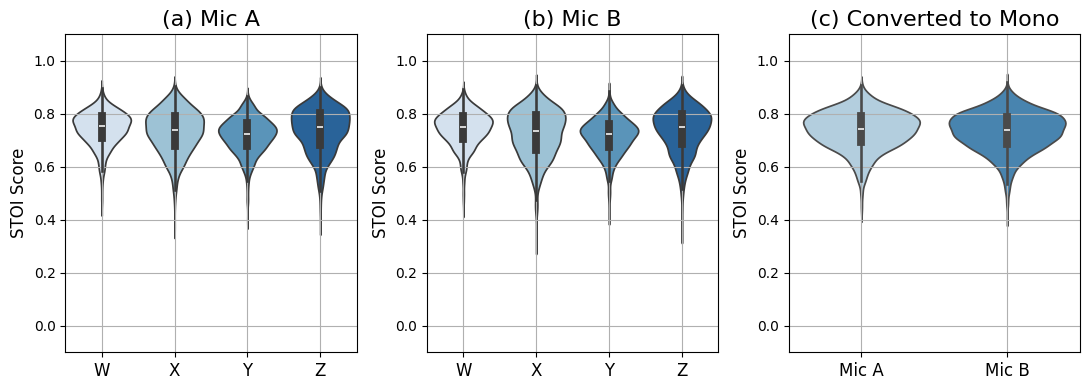}
\caption{STOI scores for the enhancement promoted by the FC U-net, using the test set for Task 1 of L3DAS22.}
\label{fig:stoifcu}
\end{figure}

Figure \ref{fig:stoifcu} illustrates violin plots for the STOI scores obtained using the FC U-net to independently enhance each channel of (a) Mic A and (b) Mic B, and (c) after conversion to Mono format in the DOA estimated by SELDnet.

\begin{figure}[ht!]
\centering
\includegraphics[width=0.75\textwidth]{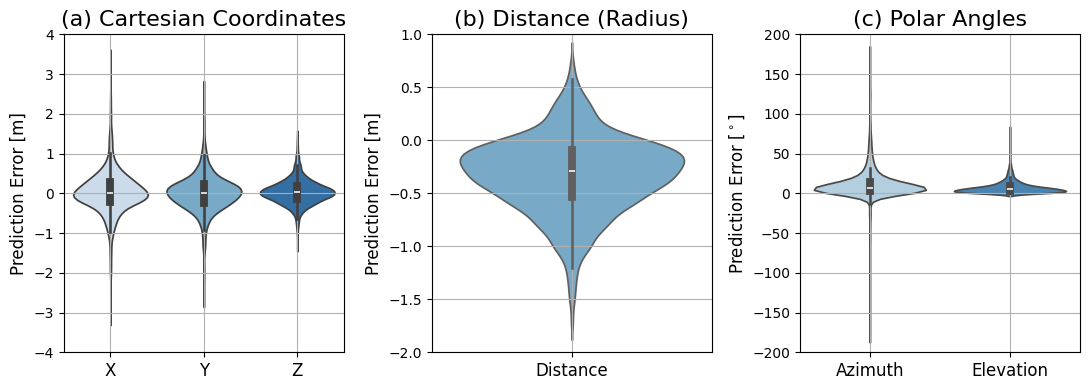}
\caption{Prediction error between SELDnet's DOA estimations using Mic A and Mic B recordings enhanced by the FC U-net as input and the test set ground-truth coordinate values in (a) Cartesian coordinates and the associated (b) distance (radius) and (c) polar angles.}
\label{fig:errorsfcu}
\end{figure}

Since SELDnet can take two sets of FOA microphones as inputs to better estimate the sound source DOAs, Figure \ref{fig:errorsfcu}(a) illustrates the prediction errors between SELDnet DOA estimations using both Mic A and Mic B recordings enhanced by the FC U-net as the input and the test set ground-truth values. Figures \ref{fig:errorsfcu}(b) and (c) illustrate the associated polar coordinates for these prediction errors.

\section{Discussion}\label{disc}

When considering the enhancement promoted by FaSNet and MMUB, higher STOI scores are obtained (cf. Figure \ref{fig:fnmmub}), owing to the incorporation of array processing, i.e., they take advantage of multiple-input data to dynamically adjust the weights of neural beamformers, which not only suppresses convolutive noise, but additive as well. In addition, because there is no need to preserve the ICLD or ICPD between the input and output signals, normalizing their outputs can also contribute to higher STOI scores by increasing the gains between clean speech estimates and noisy mixes.

In contrast, when considering the enhancement promoted by the FC U-net, it not only enhances the intelligibility of sounds (cf. Figure \ref{fig:stoifcu}), but also preserves the spatial information of the scenes to a certain extent, considering the minor errors between SELDnet predictions and ground-truth DOA coordinates (cf. Figure \ref{fig:errorsfcu}). On the lower side, this particular model treats additive noise sources as part of the scene.

That being the case, this comparison between single- and multi-channel data-driven SE and dereverberation models evinces a trade-off. Since a SISO model can enhance microphone channels while preserving ICLD and ICPD, it is reasonable to compromise the sound scene enhancement, i.e., obtaining lesser gains in intelligibility score. In this scenario, the modern research focus can switch from MISO models to SISO for augmented, virtual, and mixed reality applications. And, since the single-channel approach reduces the complexity of both problem formulation and
solution derivation, future research can concentrate on other aspects such as full-bandwidth audio enhancement, competitive noise suppression, and unsupervised learning.

Considering the state-of-the-art methods, both MISO models used in our experiments fall into this category. Also, according to \cite{dosSantos2022}, the CNN --- which is the building block of FC U-net, FaSNet, and MMUB --- was the most used type of architecture in data-driven speech
and audio enhancement studies published in 2021. However, it is also important to mention that they are based on a discriminative approach, in which, by concentrating on accurate classification between clean and noisy signals, they explicitly learn decision boundaries to separate clean speech from noisy mixes.

Contemporary SISO generative models have obtained even higher intelligibility scores by delving into the underlying distributions of training datasets, which allows for creative data generation and augmentation. Examples such as Hi-Fi GANs 1 and 2 \cite{9632770} reportedly achieved STOI scores of 0.89 and 0.92 (respectively) using the DAPS\footnote{\url{https://ccrma.stanford.edu/~gautham/Site/daps.html}} dataset. Even more recently, StoRM \cite{10180108}, a diffusion-based Stochastic Regeneration Model, reportedly achieved an Extended STOI (ESTOI, cf. \cite{7539284}) score of 0.88 using various datasets, such as WSJ0\footnote{\url{https://catalog.ldc.upenn.edu/LDC93S6A}} + CHiME\footnote{\url{https://www.chimechallenge.org/}}, REVERB-WSJ0\footnote{\url{https://github.com/sp-uhh/storm}} and VOICEBANK/DEMAND\footnote{\url{https://datashare.ed.ac.uk/handle/10283/2791}}.

\section{Conclusion}\label{conc}

This study verified that data-driven single-channel SE and dereverberation methods can be applied to multi-channel scenarios as long as they preserve ICLD and ICPD in the 3D audio captured while masking undesired noise sources. Although this approach does not leverage microphone array processing to produce an output signal, it was verified that based on DOA estimations, it minimally changes the source's position in the enhanced sound scene. Regarding intelligibility scores, both MISO systems used for comparison in our experiments achieved even higher performances. However, all models considered in this study were based on a discriminative approach. Alternative contemporary SISO models based on generative approaches have obtained state-of-the-art results. Because training single-channel models requires less training data and uses less computational resources owing to simpler model designs, modern research can focus on other challenging topics such as full-bandwidth audio enhancement, competitive noise suppression, and unsupervised learning. In summary, with current technology, there is no gain in preserving spatial information without losses in intelligibility scores. Hence, a trade-off arises depending on the type of application being considered.

\section{Acknowledgments}
This study was partially sponsored by the S\~{a}o Paulo Research Foundation (FAPESP) under grants \#2017/08120-6, \#2019/22795-1, and \#2022/16168-7. We also thank Prof. Danilo Comminiello and Riccardo Fosco Gramaccioni for their valuable discussions and suggestions.

\bibliographystyle{unsrtnat}
\bibliography{references}

\end{document}